\documentclass[twocolumn]{autart}

\newtheorem{theorem}{Theorem}
\newtheorem{definition}[theorem]{Definition}
\newtheorem{proposition}[theorem]{Proposition}

\newtheorem{lemma}[theorem]{Lemma}
\newtheorem{remark}[theorem]{Remark}

\newcommand{\R}{{\mathbb R}}

\RequirePackage{amsmath, amssymb, graphicx, url,
xspace,subfigure,braket,epstopdf}

\DeclareMathOperator*\argmin{arg\,min}

\usepackage{graphicx,subfigure}
\usepackage{epsfig} 
\usepackage{mathptmx} 
\usepackage{times} 
\usepackage{amsmath} 
\usepackage{amssymb}  
\usepackage{amsfonts}
\usepackage{euscript}
\usepackage{color}

\begin{document}

\begin{frontmatter}

\thanks[footnoteinfo]{This research has been partially supported by
the MIUR FIRB project RBFR12M3AC-Learning meets time: a new
computational approach to learning in dynamic systems and by the
Progetto di Ateneo CPDA147754/14-New statistical learning approach
for multi-agents adaptive estimation and coverage control.
This paper was
not presented at any IFAC meeting. Corresponding author Gianluigi
Pillonetto Ph. +390498277607.}

\title{The interplay between system identification and machine learning}

\author[First]{Gianluigi Pillonetto}

\address[First]{Department of Information  Engineering, University of Padova, Padova, Italy (e-mail: giapi@dei.unipd.it)}

%


\maketitle
\begin{abstract}
Learning from examples is one of the key problems in science and engineering.  
It deals with function reconstruction from a finite set of direct and noisy 
samples. \emph{Regularization in reproducing kernel Hilbert spaces} (RKHSs) 
is widely used to solve this task and includes powerful estimators 
such as regularization networks. 
Recent achievements include the proof of the 
statistical consistency of these kernel-based approaches. 
Parallel to this, many different system identification techniques have been developed
but the interaction with machine learning does not appear so strong yet.
One reason is that the RKHSs usually employed in machine learning do not embed
the information available on dynamic systems, e.g. BIBO stability.
In addition, in system identification 
the independent data assumptions routinely adopted in machine learning are 
never satisfied in practice. 
This paper provides new results which strengthen the connection between 
system identification and machine learning. Our starting point is the introduction of 
RKHSs of dynamic systems. They contain functionals over spaces defined by
system inputs and allow 
to interpret system identification as 
learning from examples.
In both linear and nonlinear settings,
it is shown that this perspective permits to derive in a relatively simple way   
conditions on RKHS stability (i.e. the property
of containing only BIBO stable systems or predictors), also facilitating the design  of
new kernels for system identification. 
Furthermore, we 
prove the convergence of the 
regularized estimator to the optimal predictor
under conditions typical of dynamic systems.
\end{abstract}

\end{frontmatter}

\section{Introduction}
Learning from examples is key in science and engineering,
considered at the core of intelligence's understanding \cite{PogShe99}.  
In mathematical terms, it can be described as follows. We are given a finite set of training data $(x_i, y_i)$,
where $x_i$ is the so called input location while $y_i$ is the corresponding output measurement. 
The goal is then the reconstruction of a function 
with good prediction capability on future data. This means that, for a new 
pair $(x, y)$, the prediction $g(x)$ should be close to $y$.\\
To solve this task, nonparametric techniques have been extensively used 
in the last years. Within this paradigm, 
instead of assigning to the unknown function a specific parametric structure,
$g$ is searched over a possibly infinite-dimensional functional space.
The modern approach uses Tikhonov regularization theory \cite{Tikhonov,Bertero1} in conjunction with
Reproducing Kernel Hilbert Spaces (RKHSs) \cite{Aronszajn50,Bergman50}.
RKHSs possess many important properties, being 
in one to one correspondence with the class of positive definite kernels.
Their connection with Gaussian processes
is also described in \cite{Kimeldorf71Bayes,Lukic,Bell2004,AravkinNN}.\\
While applications of RKHSs in statistics, approximation theory and computer vision
trace back to \cite{Bertero:1988,Wahba1990,Poggio90}, these spaces were introduced
to the machine learning community in \cite{Girosi:1997}. 
RKHSs permit to
treat in an unified way many different regularization methods.
The so called kernel-based methods 
\cite{PoggioMIT,Scholkopf01b} include smoothing splines \cite{Wahba1990}, regularization networks
\cite{Poggio90}, Gaussian regression \cite{Rasmussen}, and support vector machines \cite{Drucker97,Vapnik98}.
In particular, a regularization network (RN) has the structure
\begin{equation}\label{RN}
\hat{g} = \arg \min_{f \in {\mathcal{H}}}  \ \sum_{i=1}^N \frac{\left( y_i - f(x_i)
\right)^2}{N} + \gamma \| f \|^2_{\mathcal{H}} \qquad \text{RN}
\end{equation}
where $\mathcal{H}$ denotes a RKHS with norm $\| \cdot \|_{\mathcal{H}}$.
Thus, the function estimate minimizes an objective 
sum of two contrasting terms.
The first one is a quadratic loss which measures the adherence to experimental data.
The second term is the regularizer (the RKHS squared norm) which
restores the well-posedness and makes the solution depend continuously on the data.
Finally, the positive scalar $\gamma$ is the regularization parameter which has to
suitably trade off these two components.\\
The use of  (\ref{RN}) has significant advantages.
The choice of an appropriate RKHS,
often obtained just including function smoothness information \cite{Scholkopf01b},
and a careful tuning of $\gamma$, e.g. by the empirical Bayes approach \cite{Maritz:1989,ARAVKIN2012,Aravkin2014}, 
can well balance bias and variance.
One can thus obtain favorable mean squared error properties. 
Furthermore, even if $\mathcal{H}$
is infinite-dimensional, the solution $\hat{g}$ is always unique,
belongs to a finite-dimensional subspace and is available in closed-form.
This result comes from the representer theorem  \cite{Kimeldorf70,Scholkopf01,Argyriou:2009,ArgyriouD2014}.
Building upon the work  \cite{Wahba77}, 
many new results have been also recently obtained on the statistical consistency of (\ref{RN}).
In particular, the property of $\hat{g}$ to converge to the optimal predictor as the data set size grows to infinity
is discussed e.g. in \cite{Smale2007,Yuan2010,Wu2006,Muk2006,PoggioNature}.
This point is also related to Vapnik's concepts of generalization and
consistency \cite{Vapnik98}, see \cite{PoggioMIT} for connections among 
regularization in RKHS, statistical learning theory and the concept of $V_{\gamma}$ dimension as a
measure of function class complexity \cite{Alon:1997,EvgVg}.
The link between  consistency and well-posedness 
is instead discussed in \cite{Bousquet:2002,Muk2006,PoggioNature}.\\

Parallel to this, many system identification techniques have been developed in the last decades.
In linear contexts, the first regularized approaches 
trace back to \cite{Shiller1973,Akaike1979,Kitagawa1996},
see also  \cite{Goodwin1992,Ljung2014} where  
model error is described via a nonparametric structure.  
More recent approaches, also inspired by    
nuclear and atomic norms \cite{Chand2012}, can instead be found in \cite{liu2009,GrossmanCDC09,Mohan2010,Rojas2014,Pillonetto2016}.
In the last years, many nonparametric techniques have been proposed also for nonlinear system identification.
They exploit e.g. neural networks \cite{Tsungnan,Shun}, 
Volterra theory \cite{Franz06aunifying}, 
kernel-type estimators  \cite{Leithead2003,PilTAC2011,Bai2015} which include also weights optimization to control the mean squared error \cite{Roll2005,Bai2007,Bai2010}.
Important connections between kernel-based regularization and nonlinear system identification 
have been also obtained by the least squares support vector machines  \cite{Suykens2002,Suykens2010} 
and using Gaussian regression for state space models  \cite{Frigola2013a,Frigola2013b}.
Most of these approaches are inspired by machine learning, a fact 
not surprising since predictor estimation is at the core of the machine learning philosophy.
Indeed, a black-box relationship can be obtained through (\ref{RN}) 
using past inputs and outputs to define the input locations (regressors).
However, the kernels 
currently used for system identification 
are those conceived by the machine learning community for the reconstruction of {\it static} maps.
RKHSs suited to linear system identification, 
e.g. induced by \emph{stable spline kernels} which embed information on impulse response regularity and stability, have been proposed only recently \cite{SS2010,SS2011,ChenOL12}. Furthermore, while 
stability of a RKHS (i.e. its property of
containing only stable systems or predictors) is treated in \cite{Carmeli,SurveyKBsysid,DinuzzoSIAM15}, 
the nonlinear scenario still appears unexplored. 
Beyond stability, we also notice that the most used kernels for nonlinear regression, like the Gaussian and the Laplacian \cite{Scholkopf01b}, 
do not include other important information on dynamic systems like
the fact that output energy is expected to increase if input energy augments.\\

Another aspect that weakens the interaction between system identification and
machine learning stems also from 
the (apparently) different contexts these disciplines are applied to.
In machine learning one typically assumes that data $(x_i,y_i)$
are i.i.d. random vectors assuming values on a bounded subset of the Euclidean space. 
But in system identification, even when the system input is white noise,  
the input locations 
are not mutually independent. 
Already in the classical Gaussian noise setting, 
the outputs are not even bounded,
i.e. there is no compact set containing them with probability one.
Remarkably, this implies that 
none of the aforementioned consistency results developed for kernel-based methods can be applied.
Some extensions to the case of correlated samples can be found in \cite{LR2,LR3,LR4}  
but still under conditions far from the system identification setting.

In this paper we provide some new insights on the interplay 
between system identification and machine learning 
in a RKHS setting. Our starting point is the introduction 
of what we call \emph{RKHSs of dynamic systems} which contain 
functionals over input spaces $\mathcal{X}$ induced by system inputs $u$.
More specifically, 
each input location $x \in\mathcal{X}$ contains a piece of the trajectory of $u$ so that
any $g \in \mathcal{H}$ can be associated to a dynamic system.
When $u$ is a stationary stochastic process, its distribution
then defines the probability measure 
on $\mathcal{X}$ from which the input locations are drawn. 
Again, we stress that this framework has been 
(at least implicitly) used in previous works 
on nonlinear system identification, see e.g. 
\cite{Sjoberg:1995,PilTAC2011,Suykens,Tsungnan,Shun}.
However, it has never been cast and studied in 
its full generality under a RKHS perspective.\\ 

At first sight, our approach could appear cumbersome.  
In fact, the space $\mathcal{X}$ can turn out complex and unbounded just when the system input is Gaussian. Also, 
$\mathcal{X}$ could be a function space itself (as e.g. happens in continuous-time). 
It will be instead shown that 
this perspective is key to obtain the following achievements: 
\begin{itemize}
\item  linear and nonlinear system identification
can be treated in an unified way in both discrete- and continuous-time. 
Thus, the estimator (\ref{RN}) can be used in many different contexts, relevant for the control community, 
just changing the RKHS. This is important for the development of a general theory which links 
regularization in RKHS and system identification;
\item system input's role in determining the nature of the RKHS
is made explicit. This will be also described  in more detail in the linear system context, illustrating
the distinction between the concept of RKHSs $\mathcal{H}$ of dynamic systems and that of RKHSs $\mathcal{I}$ of 
impulse responses;
\item for linear systems 
we provide a new and simple derivation of the necessary and sufficient condition  
for RKHS stability \cite{Carmeli,SurveyKBsysid,DinuzzoSIAM15} that relies just on basic RKHS theory; 
\item in the nonlinear scenario, we obtain a sufficient condition for RKHS stability
which has wide applicability. We also derive a
new stable kernel for nonlinear system identification;
\item consistency of the RN (\ref{RN}) is proved under assumptions 
suited to system identification, revealing the link between 
consistency and RKHS stability. 
 
\end{itemize}

The paper is organized as follows. In Section \ref{Sec2} we  provide a brief
overview on RKHSs. 
In Section \ref{Sec3}, the concept of RKHSs of dynamic systems is defined
by introducing input spaces $\mathcal{X}$ induced by system inputs.
The case of linear dynamic systems is then detailed via its relationship
with linear kernels. The difference between
the concepts of RKHSs of dynamic systems and RKHSs of impulse responses
is also elucidated. 
Section \ref{Sec4}  discusses the concept of stable RKHS.
We provide a new simple characterization
of RKHS stability in the linear setting. Then, a sufficient condition
for RKHS stability is worked out  in the nonlinear scenario. 
We also introduce a new kernel 
for nonlinear system identification, testing its effectiveness on a benchmark problem. 
In Section \ref{Sec5}, we first review the connection between
the machine learning concept of \emph{regression function} and 
that of \emph{optimal predictor} encountered in system identification.
Then, the consistency of the RN
(\ref{RN}) is proved in the general framework of RKHSs of dynamic systems.
Conclusions end the paper while proofs of the consistency results are
gathered in Appendix.\\

In what follows, the analysis is always restricted to causal systems and, 
to simplify the exposition, the input locations contain only past inputs so that
output error models are considered.
If an autoregressive part is included, 
the consistency analysis in Section \ref{Sec5} remains unchanged while
the conditions developed in Section \ref{Sec4} 
guarantee predictor (in place of system) stability.

\section{Brief overview on RKHSs}\label{Sec2}

We use $\mathcal{X}$ to indicate a function domain. This is a non-empty set 
often referred to as the \emph{input space} in machine learning. Its generic element 
is the \emph{input location}, denoted by $x$ or $a$ in the sequel. 
All the functions are assumed real valued,
so that $g:\mathcal{X} \rightarrow \mathbb{R}$.\\  
In function estimation problems, the goal is 
to estimate maps to make predictions over the whole $\mathcal{X}$.
Thus, a basic requirement is to use an hypothesis space $\mathcal{H}$ with functions  
well defined pointwise for any $x \in\mathcal{X}$. 
In particular, assume that all the pointwise evaluators $g \rightarrow g(x)$ 
are linear and bounded over $\mathcal{H}$, i.e. $\forall x \in\mathcal{X}$ 
there exists $C_x< \infty$ such that 
\begin{equation}\label{RKHS}
|g(x)| \leq C_x\|g\|_{\mathcal{H}}, \quad \forall g \in \mathcal{H}.
\end{equation}
This property already leads to the spaces of interest.

\begin{definition}[RKHS]\label{DefRKHS}
A reproducing kernel Hilbert space (RKHS) $\mathcal{H} $over $\mathcal{X}$ is a Hilbert space 
containing functions $g:\mathcal{X} \rightarrow \mathbb{R}$ where (\ref{RKHS}) holds.
\end{definition}

RKHSs are  
connected to the concept of positive definite kernel,
a particular function defined over $\mathcal{X}\times\mathcal{X}$.

\begin{definition}[Positive definite kernel and kernel section]\label{ch01-D03}
A symmetric function $K :\mathcal{X}\times\mathcal{X} \rightarrow \mathbb{R}$ is called \emph{positive definite kernel} if, for any integer $p$, it holds
$$
\sum_{i=1}^{p}\sum_{j=1}^{p}c_ic_j \mathcal{K}(x_i,x_j) \geq 0, \quad \forall (x_k,c_k) \in \left(\mathcal{X},\mathbb{R}\right), \quad k=1,\ldots, p.
$$
The \emph{kernel section} $\mathcal{K}_x$ centered at $x$ is the function
from $\mathcal{X}$ to $\mathbb{R}$ defined by 
$$
\mathcal{K}_x(a) = \mathcal{K}(a,x) \quad  \forall a \in\mathcal{X}.
$$
\end{definition}

The following theorem provides the
one-to-one correspondence between RKHSs and positive definite kernels.

\begin{theorem}[Moore-Aronszajn and reproducing property]\label{Math}
To every RKHS $\mathcal{H}$ there corresponds a unique positive definite kernel $\mathcal{K}$ such that the 
so called \emph{reproducing property} holds, i.e.
\begin{equation}\label{RepProp}
\langle \mathcal{K}_x, g \rangle = g(x) \quad   \forall (x, g) \in \left(\mathcal{X},\mathcal{H}\right)
\end{equation} 
Conversely, given a positive definite kernel $\mathcal{K}$, there exists a unique RKHS of real-valued functions defined 
over $\mathcal{X}$ where (\ref{RepProp}) holds. 
\end{theorem}

Theorem \ref{Math} shows that a RKHS $\mathcal{H}$ is completely defined by
a kernel $\mathcal{K}$, also called the \emph{reproducing kernel} of $\mathcal{H}$.
More specifically, it can be proved that any RKHS is generated by the kernel sections in the following manner.
Let $S$ denote the subspace spanned by $ \{ \mathcal{K}_x \}_{ x \in\mathcal{X} }$ and
for any $g \in S$, say $g = \sum_{i=1}^{p} c_i \mathcal{K}_{x_i}$,
define the norm 
\begin{equation}\label{IndNorm}
\|g \|_{\mathcal{H}}^2 =  \sum_{i=1}^p \sum_{j=1}^p c_i c_j \mathcal{K}(x_i,x_j).
\end{equation}
Then, one has that $\mathcal{H}$ is the union of $S$
and all the limits w.r.t. $\| \cdot \|_{\mathcal{H}}$ of the Cauchy sequences contained in $S$.
A consequence of this construction is that any $g \in \mathcal{H}$ inherits kernel properties, e.g.
continuity of $\mathcal{K}$ implies that all the $g \in \mathcal{H}$ are continuous \cite{Cucker01}[p. 35].\\
The kernel sections play a key role also in providing 
the closed-form solution of the RN (\ref{RN}),
as illustrated in the famous representer theorem.

\begin{theorem}[Representer theorem]\label{THM01}
The solution of (\ref{RN}) is unique and given by 
\begin{equation}\label{GeneralSol}
\hat{g} = \sum_{i=1}^N \ \hat{c}_i \mathcal{K}_{x_i},
\end{equation}
where the scalars $\hat{c}_i$ are the components of the vector
\begin{equation}\label{RNcoeff}
\hat{c} = \left(\mathbf{\mathbf{K}}+\gamma N  I_{N}\right)^{-1}Y,
\end{equation}
$Y$ is the column vector with $i$-th element $y_i$,
$I_{N}$ is the $N \times N$ identity matrix and the $(i,j)$ entry of $\mathbf{K}$ is $\mathcal{K}(x_i,x_j)$.
\end{theorem}


Another RKHS characterization useful in what follows
is obtained when the kernel can be diagonalized as follows 
\begin{equation}\label{RKHSdepfun}
\mathcal{K}(a,x) = \sum_{i=1}^{\infty} \ \zeta_i \rho_i(a) \rho_i(x), \ \ \zeta_i > 0 \ \forall i. 
\end{equation}
The RKHS is then separable and the following result holds, e.g. see \cite{PoggioMIT}[p. 15] and  \cite{Cucker01}[p. 36].
\begin{theorem}[Spectral representation of a RKHS] \label{SpectralRKHS3} 
Let (\ref{RKHSdepfun}) hold and assume that the $\rho_i$ form a set of linearly independent functions on $\mathcal{X}$.
Then, one has
\begin{equation}\label{Hrep3}
\mathcal{H}  =  \left\{ g   \ | \   g(x) = \sum_{i=1}^{\infty} c_i \rho_i(x) \ \mbox{s.t.} \  
\sum_{i=1}^{\infty}  \frac{c_i^2}{\zeta_i } < \infty
\right\},
\end{equation}
and
\begin{equation}\label{Hrep5}
\langle f,g \rangle_{\mathcal{H}} =  \sum_{i=1}^{\infty} \frac{b_i c_i}{\zeta_i}, \quad \| f\|_{\mathcal{H}}^2 =  \sum_{i=1}^{\infty} \frac{b_i^2}{\zeta_i},
\end{equation}
where $f = \sum_{i=1}^{\infty} b_i \rho_i$ and $g = \sum_{i=1}^{\infty} c_i \rho_i$.
\end{theorem}
The expansion (\ref{RKHSdepfun}) can e.g. be obtained by the Mercer theorem \cite{Mercer1909,Hochstadt73}.
In particular, let $\mu_x$ be a nondegenerate $\sigma$-finite measure on $\mathcal{X}$. 
Then, under somewhat general conditions \cite{Sun05}, 
the $\rho_i$ and $\zeta_i$ in (\ref{RKHSdepfun}) can be set to the eigenfunctions and eigenvalues
of the integral operator induced by $\mathcal{K}$, i.e.
\begin{equation}\label{Lk2}
\int_{\mathcal{X}} \mathcal{K}(\cdot,x) \rho_i(x) d\mu_x(x) = \zeta_i \rho_i(\cdot), \quad 0 < \zeta_1 \leq \zeta_2 \leq \ \ldots
\end{equation}
In addition, the $\rho_i$ form a complete orthonormal basis in the classical Lebesgue space 
$\mathcal{L}_2^{\mu_x}$ of functions square integrable  under $\mu_x$.\footnote{Thus, the representation (\ref{Hrep3}) is not unique since spectral maps are 
not unique.  Eigendecompositions depend on the measure $\mu_x$ but 
lead to the same RKHS.  \\}


\section{RKHSs of dynamic systems and the linear system scenario} \label{Sec3}

\subsection{RKHSs of dynamic systems}

The definition of \emph{RKHSs of dynamic systems} given below relies on simple constructions of
input spaces $\mathcal{X}$ induced by system inputs $u$.

\paragraph*{Discrete-time} First, the discrete-time setting is considered. 
Assume we are given a system input $u: \mathcal{Z} \rightarrow \mathbb{R}$.
Then, we think of any input location in $\mathcal{X}$  indexed by the time $t \in \mathcal{Z}$. 
Different cases arise depending on the postulated system model. 
For example, one can have
\begin{equation}\label{xsysid}
x_t = [u_t \ \ u_{t-1} \ \ \ldots  \ \ u_{t-m+1}]^T,
\end{equation} 
where $m$ is the system memory. 
This construction is connected to 
FIR or NFIR models and makes  $\mathcal{X}$ a subset of the 
classical Euclidean space $\mathbb{R}^m$.\\
Another scenario is 
\begin{equation}\label{xsysid2}
x_t = [u_t \ \ u_{t-1} \  \ u_{t-2}   \ \  \ldots ]^T,
\end{equation} 
where any input location is a sequence (an infinite-dimensional column vector)
and the input space $\mathcal{X}$ becomes a subset of $\mathbb{R}^{\infty}$. 
The definition (\ref{xsysid2}) is related to infinite memory systems, e.g. IIR models in linear settings.\\

\paragraph*{Continuous-time} The continuous-time
input is the map
$u: \mathbb{R} \rightarrow \mathbb{R}$. In this case, 
the input location $x_t $ becomes the function $x_t : \mathbb{R}_+ \rightarrow \mathbb{R}$
defined by 
\begin{equation}\label{xsysid3}
x_t (\tau) = u(t-\tau), \quad \tau \geq 0,
\end{equation}
i.e. $x_t$ contains the input's past up to the instant $t$.
In many circumstances, one can assume $\mathcal{X} \subset \mathcal{P}^c$,
where $\mathcal{P}^c$ contains piecewise continuous functions on $\mathbb{R}_+$.
When the input is causal, and $u_t$ is smooth for $t \geq 0$, 
the $x_t $ is indeed piecewise continuous.\\ 
Note that (\ref{xsysid3}) is the continuous-time counterpart of (\ref{xsysid2}) while 
that of (\ref{xsysid}) can be obtained just zeroing part of the input location, i.e.
\begin{equation}\label{xsysid4}
x_t (\tau) = u(t-\tau)\xi_T(\tau), \quad \tau \geq 0,
\end{equation}
where $\xi_T$ is the indicator function of the interval $[0,T]$. 
In linear systems,
(\ref{xsysid4}) arises when the impulse response support is compact.\\


RKHSs $\mathcal{H}$ of functions over domains $\mathcal{X}$, induced by system inputs $u$ as
illustrated above, are hereby called \emph{RKHSs of dynamic systems}. 
Thus, if $g \in \mathcal{H}$, the scalar $g(x_t)$ is the noiseless output at $t$ of the system 
fed with the input trajectory contained in $x_t $. 
Note that $g$ 
in general is a functional: in the cases (\ref{xsysid2}-\ref{xsysid4}) the arguments $x_t $ 
entering $g(\cdot)$ are infinite-dimensional objects.

\subsection{The linear system scenario}\label{Sec3.2}

RKHSs of linear dynamic systems
are now introduced also discussing
the structure of the resulting RN.\\
Linear system identification
was faced In \cite{SS2010} and \cite{SurveyKBsysid}[Part III] by introducing \emph{RKHSs of impulse responses}.
These are spaces $\mathcal{I}$
induced by kernels $K$ defined over subsets of $\mathbb{R}_+ \times \mathbb{R}_+$.
They thus contain causal functions, 
each of them representing an impulse response $\theta$.
The RN which returns the impulse response estimate was
\begin{equation}\label{RN2}
\hat{\theta} = \arg \min_{\theta \in {\mathcal{I}}}  \ \sum_{i=1}^N \frac{\left( y_i - (\theta \otimes u)_{t_i}
\right)^2}{N} + \gamma \| \theta \|^2_{\mathcal{I}},
\end{equation}
where $(\theta \otimes u)_{t_i}$ is the convolution between the impulse response 
and the input evaluated at $t_i$.\\
The \emph{RKHSs of linear dynamic systems} here introduced 
are instead associated to (output) linear kernels
$\mathcal{K}$ defined on $\mathcal{X} \times \mathcal{X}$ 
through convolutions of $K$ with system inputs.
In particular, if $x_t$ and $x_{\tau}$ are as in (\ref{xsysid}-\ref{xsysid4}), one has\footnote{Translated in a stochastic setting, the (output) kernel $\mathcal{K}$ 
can be seen as the covariance of a 
causal random process of covariance $K$ filtered by $u$.} 
$$
\mathcal{K}(x_t,x_{\tau}) =  \left( u \otimes (K \otimes u)_{\tau} \right)_t
$$
which e.g. in continuous-time becomes
\begin{equation}\label{ExCT}
\mathcal{K}(x_t,x_{\tau}) = \int_{0}^{+\infty} u(t-\alpha) \left( \int_{0}^{+\infty} u(\tau-\beta)  K(\alpha,\beta) d\beta \right) d\alpha.
\end{equation}
These kernels lead to the RN (\ref{RN})
which corresponds to  (\ref{RN2}) after the ``reparametrization"  
$g(x_t)=(\theta \otimes u)_t$ so that, in place of the impulse response $\theta$,
the optimization variable becomes 
the functional $g(\cdot)$.\\
The kernels $\mathcal{K}$ arising in discrete- and continuous-time are 
described below in (\ref{FIRker},\ref{IIRker}) and 
(\ref{CTker}). The distinction between the RKHSs induced by
$K$ and $\mathcal{K}$ will be further discussed in Section \ref{KvsP}. 
\vspace{-0.7cm}

\paragraph*{FIR models} We start assuming that the input location is defined by
(\ref{xsysid}) so that any $x_t $ is an $m$-dimensional (column) vector 
and $\mathcal{X} \subseteq \R^m$. 
If $K \in \R^{m \times m}$ is a symmetric and positive semidefinite matrix,
a linear kernel is defined as follows
\begin{equation}\label{FIRker}
\mathcal{K}(a,x) = a^T K x, \quad (a,x) \in \mathbb{R}^m \times \mathbb{R}^m.
\end{equation}
All the kernel sections are linear functions. Their span defines a finite-dimensional (closed)
subspace that, in view of the discussion following Theorem \ref{Math}, 
coincides with the whole $\mathcal{H}$.
Hence, $\mathcal{H}$  
is  a space of linear functions: for any $g \in \mathcal{H}$, 
there exists $a \in \R^m$ such that 
$$
g(x)=a^T K x=\mathcal{K}_a(x).
$$
If $K$ is full rank, it holds that
\begin{eqnarray*}\nonumber
\qquad \qquad \qquad || g ||^2_{\mathcal{H}}  &=&   || \mathcal{K}_a ||^2_{\mathcal{H}} = \langle \mathcal{K}_a, \mathcal{K}_a \rangle_{\mathcal{H}} \\ 
\qquad  \qquad \qquad &=& \mathcal{K}(a,a) = a^T K a \\ 
\qquad \qquad \qquad &=&  \theta^T K^{-1} \theta \ \   \textit{with} \ \  \theta := K a.
\end{eqnarray*}
Let us use the $\mathcal{H}$ associated to (\ref{FIRker}) as hypothesis space for the RN 
in (\ref{RN}). 
Let $Y=[y_1 \ \ldots y_N]^T$ and $\Phi \in \mathbb{R}^{N \times m}$ with $i$-th row equal to
$x_i^T$, where 
$$
x_i:=x_{t_i} \quad y_i:=y_{t_i},
$$   
and $t_i$ is the time instant where the $i$-th output is measured.
Then, after plugging the representation $g(x)=\theta^T x$ in (\ref{RN}),
one obtains 
$\hat{g}(x)=\hat{\theta}^Tx$ with  
\begin{subequations} \label{RKHSlinreg}
\begin{align} \label{RKHSlinreg1}
\hat{\theta} &= \argmin_{\theta \in \mathbb{R}^{m}} \ \|Y-\Phi\theta\|^2+ \gamma \theta^T K^{-1} \theta \\
&= (\Phi^T \Phi + \gamma P^{-1})^{-1} \Phi^T Y.  \label{RKHSlinreg2}
\end{align}
\end{subequations}
The nature of the input locations (\ref{xsysid}) shows that  
$\hat{\theta}$ is the impulse response estimate.
Thus,  (\ref{RKHSlinreg}) corresponds to regularized 
FIR estimation as e.g. discussed in \cite{ChenOL12}.\\  
\vspace{-1cm}


\paragraph*{IIR models}
Consider now the input locations defined by
(\ref{xsysid2}). The input space contains sequences
and $\mathcal{X} \subseteq \mathbb{R}^\infty$.
Interpreting any input location as an infinite-dimensional
column, we can use ordinary algebra's notation 
to handle infinite-dimensional objects. For example, 
if $(a,x) \in (\mathcal{X},\mathcal{X})$ then $a^Tx=\langle a,x \rangle_2$,
where $\langle \cdot,\cdot \rangle_2$ is the inner-product in the classical space
$\ell_2$ of squared summable sequences.\\
Let $K$ be symmetric and positive semidefinite  
infinite-dimensional matrix $K$ (the nature of $K$ is discussed also in Section \ref{KvsP}). 
Then, the function  
\begin{equation}\label{IIRker}
\mathcal{K}(x,a) = x^T K a, \quad (x,a) \in \mathbb{R}^{\infty} \times \mathbb{R}^{\infty}
\end{equation}
defines a linear kernel on $\mathcal{X} \times\mathcal{X}$.
Following arguments similar to those developed in the FIR case, 
one can see that the RKHS associated to such $\mathcal{K}$ contains linear functions of the form $g(x)=a^T K x$
with $a \in \mathbb{R}^{\infty}$. Note that each $g \in \mathcal{H}$ 
is a functional defined by the sequence $a^T K$ which represents an impulse response.
In fact, 
one can deduce from (\ref{xsysid2}) that $g(x_t)$ is the discrete-time convolution, evaluated at $t$, 
between $u$ and $a^T K$.\\
The RN with $\mathcal{H}$ induced by (\ref{IIRker})
now implements regularized IIR estimation.
Roughly speaking, 
(\ref{RN}) becomes the limit of (\ref{RKHSlinreg}) 
for $m \rightarrow \infty$. The exact solution can be obtained by   
the representer theorem
(\ref{GeneralSol}) and turns out 
\begin{equation} \label{IIRsol}
\hat{g}(x) = \sum_{i=1}^N \ \hat{c}_i \mathcal{K}_{x_i}(x) = \hat{\theta}^T x,
\end{equation}
where the $\hat{c}_i$ are the components of (\ref{RNcoeff}) while the infinite-dimensional column vector
\begin{equation} \label{IIRsol2}
\hat{\theta}  := \sum_{i=1}^N \ \hat{c}_i K x_i
\end{equation}
contains the impulse response coefficients estimates.\\
\vspace{-1cm}


\paragraph*{Continuous-time} 

The continuous-time scenario arises 
considering the input locations defined by
(\ref{xsysid3}) or (\ref{xsysid4}). The input space $\mathcal{X}$ 
now contains causal functions.
Considering (\ref{xsysid3}), 
given a positive-definite kernel $K: \mathbb{R}_+ \times \mathbb{R}_+ \rightarrow \mathbb{R}$, 
the linear kernel $\mathcal{K}$ is
\begin{equation}\label{CTker}
\mathcal{K}(x,a) =  \int_{\mathbb{R}_+ \times \mathbb{R}_+}    \ K(t,\tau) x(t)a(\tau) dt d\tau
\end{equation}
which coincides with (\ref{ExCT}) when $x=x_t$ and $a=x_{\tau}$.
Each kernel section $\mathcal{K}_x(\cdot)$ is a continuous-time linear system  
with impulse response $\theta(\cdot)=\int_{\mathbb{R}_+}    \ K(\cdot,t) x(t) dt$.
Thus, the corresponding RKHS contains linear functionals and 
(\ref{RN}) now implements regularized system identification in continuous-time.
Using the representer theorem, the solution of (\ref{RN}) is 
\begin{equation} \label{CTsol}
\hat{g}(x) = \sum_{i=1}^N \ \hat{c}_i \mathcal{K}_{x_i}(x) = \int_{\mathbb{R}_+} \hat{\theta}(\tau) x(\tau) d \tau 
\end{equation} 
where $\hat{c}$ is still defined by (\ref{RNcoeff}) while
$\hat{\theta}$  is the impulse response estimate given by
\begin{equation} \label{CTsol2}
\hat{\theta}(\tau)  := \sum_{i=1}^N \ \hat{c}_i   \int_{\mathbb{R}_+}  \ K(\tau,t) x_i(t) dt.
\end{equation} 


\subsection{Relationship between RKHSs of impulse responses and RKHSs of dynamic systems}\label{KvsP}  

In (\ref{IIRker}), the infinite-dimensional matrix $K$ represents a 
kernel over $\mathbb{N} \times \mathbb{N}$. 
Then, let $\mathcal{I}$ be the corresponding RKHS which contains 
infinite-dimensional column vectors $\theta=[\theta_1 \ \theta_2 \ldots ]^T$.
We will now see that $\mathcal{I}$ is the RKHS of impulse responses associated 
to $\mathcal{H}$, i.e. each $\theta \in \mathcal{I}$ is the impulse response
of a linear system $g \in \mathcal{H}$. 
In particular, let $K$ 
admit the following expansion in terms of linearly independent infinite-dimensional 
(column) vectors $\psi_i$:
$$
K= \sum_{i=1}^{\infty} \zeta_i \psi_i \psi_i^T.
$$
According to Theorem \ref{SpectralRKHS3}, the span of the $\psi_i$ 
provides all the $\theta \in  \mathcal{I}$. Moreover, if  $\theta = \sum_{i=1}^{\infty} c_i \psi_i$, then 
$
\| \theta\|_{\mathcal{I}}^2 = \sum_{i=1}^{\infty} \frac{c_i^2}{\zeta_i}.
$
The equality
\begin{subequations}
\begin{align*}
\mathcal{K}(a,x) &= a^TKx = a^T \left( \sum_{i=1}^{\infty} \zeta_i \psi_i \psi_i^T \right)  x\\
&=   \sum_{i=1}^{\infty} \zeta_i 
\underbrace{\left(  a^T \psi_i \right)}_{\rho_i(a)} \underbrace{ \left( \psi_i^T x  \right)}_{\rho_i(x)},
\end{align*}
\end{subequations}
also provides the expansion of $\mathcal{K}$
in terms of functionals $\rho_i(\cdot)$
defined by $\rho_i(x):=\psi_i^T x$.
Assuming that such functionals are linearly independent,
it comes from Theorem \ref{SpectralRKHS3}  that
each dynamic system $g \in \mathcal{H}$ has the representation 
$g(\cdot) = \sum_{i=1}^{\infty} c_i \rho_i(\cdot)$. It is now obvious that 
such system is associated to the
impulse response $\theta = \sum_{i=1}^{\infty} c_i \psi_i$ and the two spaces
are isometrically isomorphic since
$$
\|g\|_{\mathcal{H}}^2 = \sum_{i=1}^{\infty} \frac{c_i^2}{\zeta_i}=\| \theta\|_{\mathcal{I}}^2.
$$ 
This result holds also in continuous-time
where $\mathcal{I}$ is now the RKHS associated to the kernel 
$K: \mathbb{R}_+ \times \mathbb{R}_+ \rightarrow \mathbb{R}$. 
Letting $\psi_i$ be real-valued functions on $\mathbb{R}_+$, 
this comes from the same arguments adopted in discrete-time 
but now applied to the expansions
$$
K(t,\tau)= \sum_{i=1}^{\infty} \zeta_i \psi_i(t) \psi_i(\tau),
$$
and
\begin{subequations}
\begin{align*}
\mathcal{K}(x,a) & =  \int_{\mathbb{R}_+ \times \mathbb{R}_+}    \ K(t,\tau) x(t)a(\tau) dt d\tau \\
& =  \sum_{i=1}^{\infty}   \zeta_i  
\underbrace{\left( \int_{\mathbb{R}_+}   \psi_i(t) x(t) dt \right)}_{\rho_i(x)}
\underbrace{\left(\int_{\mathbb{R}_+}   \psi_i(\tau) a(\tau) d\tau \right)}_{\rho_i(a)}.
\end{align*}
\end{subequations}
 
 \begin{remark}\label{NonIsometry}
The functionals $\rho_i$ could turn out linearly dependent even if the $\psi_i$ 
composing $K$ are linearly independent.
This depends on the nature of the input space. 
For instance, let $\mathcal{X}$ contain only the input locations
induced by $u_t=\sin(\omega t)$. Then, if $\psi_i$ is a rational transfer function
with zeros $\pm j \omega$, the functional $\rho_i$  
associated to $\psi_i$ vanishes over $\mathcal{X}$. 
In these cases, there is no isometry between $\mathcal{H}$  and $\mathcal{I}$:
the same dynamic system  $g$
could be defined by different impulse responses $\theta^i \in \mathcal{I}$.
In particular, results on 
RKHSs induced by sums of kernels reported in \cite{Aronszajn50}[Section 6 on p. 352] 
allow us to conclude that $\|g\|_{\mathcal{H}}=\min_i  \|\theta^i\|_{\mathcal{I}}$.
Thus, among all the possible equivalent representations in $\mathcal{I}$ of the dynamic system
$g \in \mathcal{H}$, the complexity of $g$ is quantified by that of minimum norm.
This is illustrated through the following simple continuous-time example. 
Assume that
$$
K(t,\tau)= \zeta_1  \psi_1(t) \psi_1(\tau)  +  \zeta_2 \psi_2(t) \psi_2(\tau), 
$$ 
where the Laplace transforms of $\psi_1$ and $\psi_2$ 
are given, respectively, by the rational transfer functions
$$
W_1(s)=\frac{2s}{s+1+\sqrt{2}}, \quad W_2(s)=\frac{s+1-\sqrt{2}}{s+1},
$$
which satisfy $W_1(\sqrt{-1})=W_2(\sqrt{-1})$.
Let the input space contain only the input locations
induced by $u_t=\sin(t)$. 
Then, the functionals $(\rho_1,\rho_2)$ associated, respectively, to $(\psi_1,\psi_2)$
coincide over the entire $\mathcal{X}$. Thus, the two impulse responses $\psi_1$ and $\psi_2$
induce the same system $g \in \mathcal{H}$. 
Using Theorem \ref{SpectralRKHS3}, one has
$$
\| \psi_1 \|^2_{\mathcal{I}} = \frac{1}{\zeta_1}, \quad  \| \psi_2 \|^2_{\mathcal{I}} = \frac{1}{\zeta_2},
$$ 
which implies
$$
\| g\|^2_{\mathcal{H}} = \min\left(\frac{1}{\zeta_1},\frac{1}{\zeta_2}\right).
$$

\end{remark}

\section{Stable RKHSs} \label{Sec4}

BIBO stability of a dynamic system is a familiar notion in control.
In the RKHS context, we have the following definition.

\begin{definition}[stable dynamic system] \label{DefDynStab}
Let $u$ be any bounded input, i.e. satisfying $|u_t|<M_u< \infty \ \forall t$.
Then, the dynamic system $g \in \mathcal{H}$ is said to be (BIBO) stable if
there exits a constant $M_y<\infty$ such that $|g(x_t)|<M_y$ for any $t$ and any 
input location $x_t$ induced by $u$.
\end{definition}
Note that, for $g \in \mathcal{H}$ to be stable, the above definition implicitly requires 
the input space of $\mathcal{H}$ to contain any $x_t$ induced by any bounded input. 

\begin{definition}[stable RKHS] \label{DefStab}
Let $\mathcal{H}$ be a RKHS of dynamic systems induced by the kernel $\mathcal{K}$.
Then, $\mathcal{H}$ and $\mathcal{K}$ are said to be stable if each $g \in \mathcal{H}$ 
is stable.
\end{definition}

To derive stability conditions on the kernel, let us first introduce some 
useful Banach spaces. 
The first two regard the discrete-time setting:
\begin{itemize}
\item the space $\ell_1$ of absolutely summable real sequences $a=[a_1 \ a_2 \ldots]$, i.e. such that
$\sum_{i=1}^\infty |a_i| < \infty,$
equipped with the norm 
$$
\| a \|_1 = \sum_{i=1}^\infty |a_i|;
$$
\item the space $\ell_{\infty}$ of bounded real sequences $a=[a_1 \ a_2 \ldots]$, i.e. such that
$\sup_i |a_i| < \infty$,
equipped with the norm 
$$
\| a \|_{\infty} = \sup_i |a_i|.
$$
\end{itemize}
The other two are concerned with continuous-time: 
\begin{itemize}
\item the Lebesgue space $\mathcal{L}_1$ of functions 
$a: \mathbb{R}_+ \rightarrow \mathbb{R}$ absolutely integrable, i.e. such that
$\int_{\mathbb{R}_+} |a(t)| dt < \infty$,
equipped with the norm 
$$
\|a \|_1 = \int_{\mathbb{R}_+} |a(t)| dt;
$$
\item the Lebesgue space $\mathcal{L}_\infty$ of functions 
$a: \mathbb{R}_+ \rightarrow \mathbb{R}$
essentially bounded, i.e. for any $a$ there exists $M_a$ such that 
$$
|a(t)| \leq M_a \ \mbox{almost everywhere in $\mathbb{R}_+$}, 
$$
equipped with the norm
$$ 
\|a \|_\infty = \inf \left\{ M \ \text{s.t.} \ |a(t)| \leq M \ \mbox{a.e.} \right\} .
$$
\end{itemize}

\subsection{The linear system scenario}\label{StableLinKer}

We start studying the stability of RKHSs of linear dynamic systems.
Obviously, all the FIR kernels (\ref{FIRker}) 
induce stable RKHSs. As for the IIR and continuous-time kernels in (\ref{IIRker}) 
and (\ref{CTker}), first it is useful to 
recall the classical result linking BIBO stability and
impulse response summability.

\begin{proposition}{\bf{(BIBO stability and impulse response summability)}}\label{SumIR} 
Let $\theta$ be the impulse response of a linear system.
Then, the system is BIBO stable iff $\theta \in \ell_1$ in discrete-time or
$\theta \in \mathcal{L}_1$ in continuous-time. 
\end{proposition}

The next proposition provides the necessary and sufficient condition
for RKHS stability in the linear scenario.\\ 

\begin{proposition}[RKHS stability in the linear case]\label{StableLinear}
Let $\mathcal{H}$ be the RKHS of dynamic systems
$g:\mathcal{X} \rightarrow \mathbb{R}$ induced by the IIR kernel (\ref{IIRker}).
Then, the following statements are equivalent
\begin{enumerate}
\item $\mathcal{H}$  is stable;
\item The input space $\mathcal{X}$ contains $\ell_{\infty}$ so that 
$$
g(a)<\infty \ \   \text{for any} \  \ (g,a) \in (\mathcal{H},\ell_{\infty});
$$
\item $\sum_{i=1}^\infty \left|  \sum_{j=1}^\infty K(i,j) a_j  \right| < \infty  \ $  for any $ \ a \in
\ell_{\infty}.$
\end{enumerate}
Let instead $\mathcal{H}$ be the RKHS induced by the continuous-time kernel (\ref{CTker}).
The following statements are then equivalent
\begin{enumerate}
\item $\mathcal{H}$  is stable;
\item The input space $\mathcal{X}$ contains $\mathcal{L}_{\infty}$ so that 
$$
g(a)<\infty \  \  \text{for any}  \ \  (g,a) \in (\mathcal{H}, \mathcal{L}_{\infty});
$$
\item  $ \int_{ \mathbb{R}_+} \left| \int_{ \mathbb{R}_+}  K(t,\tau) a(\tau) d\tau \right| dt < +\infty \ $  for any $ \ a \in 
\mathcal{L}_{\infty}.$ 
\end{enumerate} 
\end{proposition}
{\bf{Proof:}} The proof is developed in discrete-time. The continuous-time case follows 
exactly by the same arguments with minor modifications.\\ 
$(1) \rightarrow (2)$ Recalling Definition \ref{DefDynStab} and subsequent discussion, 
this is a direct consequence of the BIBO stability assumption of any $g \in \mathcal{H}$.\\
$(2) \rightarrow (1)$ Given any $g \in \mathcal{H}$, let  $\theta=[\theta_1 \ \theta_2 \ \ldots]^T$ its associated impulse response and define  
\begin{equation}\label{xSign}
x_t=[ \text{sign}(\theta_1) \ \text{sign}(\theta_2) \ \ldots]^T.
\end{equation}
The assumption $g(x_t)=\theta^T x_t = \|\theta \|_1 <\infty$ implies that 
$\theta \in \ell_1$ and the implication follows by Proposition \ref{SumIR}.\\
$(2) \rightarrow (3)$ By assumption, the kernel is well defined over the entire
$\ell_{\infty} \times \ell_{\infty}$. Hence, any kernel section $\mathcal{K}_a$ centred on $a \in \ell_{\infty}$ 
is a well defined element in $\mathcal{H}$ and corresponds to a dynamic system 
with associated impulse response $\theta=Ka$. 
With $x_t$ still defined by (\ref{xSign}), one has $\mathcal{K}_a(x_t)= \|\theta \|_1<\infty$ 
which implies  
$Ka \in \ell_1$ and proves (3).\\
$(3) \rightarrow (2)$  By assumption, any impulse response
associated to any kernel section centred on $a \in \ell_{\infty}$ belongs to $\ell_1$. 
This implies 
that the kernel 
$\mathcal{K}$ associated to $\mathcal{H}$ is well defined over the entire
$\ell_{\infty} \times \ell_{\infty}$. Recalling Definition \ref{DefRKHS} and eq. (\ref{RKHS}), 
RKHS theory then ensures that any $g \in \mathcal{H}$ 
is well defined pointwise on $\ell_{\infty}$ and $g(a)<\infty \ \forall a \in \ell_{\infty}$.\\
\begin{flushright}
$\blacksquare$
\end{flushright}

Point (3) contained in Proposition \ref{StableLinear} was 
also cited in \cite{SurveyKBsysid,DinuzzoSIAM15}
as a particularization of a quite involved and abstract result 
reported in \cite{Carmeli}. The stability proof reported below turns instead out
surprisingly simple. The reason is that, with the notation
adopted in (\ref{IIRker}) and (\ref{CTker}), the outcomes in
\cite{Carmeli} were obtained starting from spaces $\mathcal{I}$ of impulse responses induced
by $K$. 
Our starting point is instead the RKHSs $\mathcal{H}$ of dynamic systems 
induced by $\mathcal{K}$ (in turn defined by $K$). This different perspective permits
to greatly simplify the analysis: 
kernel stability can be characterized just combining basic 
RKHS theory and 
Proposition \ref{SumIR}.

Proposition \ref{StableLinear} shows that   RKHS stability is implied by 
the absolute integrability of $K$, i.e. by
\begin{equation}\label{SuffLin}
\sum_{i=1}^\infty  \sum_{j=1}^\infty  |K(i,j)| < \infty \ \  \text{or}  \ \   \int_{\mathbb{R}_+ \times \mathbb{R}_+} |K(t,\tau)| dt d\tau < \infty
\end{equation}
in discrete- and continuous-time, respectively.
The condition (\ref{SuffLin}) is 
also necessary for nonnegative-valued  kernels
\cite{SurveyKBsysid}[Section 13].
Then, considering e.g. the continuous-time setting,
the popular Gaussian and Laplacian kernels,
which belong to 
the class of radial basis kernels  
$K(t,s) = h(|s-t|)$ for $t,s \geq 0$, are all unstable.
Stability instead holds 
for the  
stable spline kernel \cite{PillACC2010} given by:
\begin{equation}\label{SS1}
K(t,s) = e^{-\beta \max(t,s)} \quad t,s \geq 0 
\end{equation}
where $\beta>0$ is related to the impulse response's dominant pole.

\subsection{The nonlinear system scenario}

Let us now consider RKHSs of nonlinear dynamic systems with input locations
(\ref{xsysid}-\ref{xsysid4}). A very simple sufficient condition for RKHS stability is reported below.

\begin{proposition}[RKHS stability in the nonlinear case]\label{StableNonLinear}
Let $\mathcal{H}$ be a RKHS of dynamic systems 
induced by the kernel $\mathcal{K}$.
Let $B_{\infty}^r$ denote the closed ball of radius $r$ induced by $\| \cdot \|_{\infty}$, 
contained in $\ell_{\infty}$ or $\mathbb{R}^m$ in
discrete-time, or in $\mathcal{L}_{\infty}$ in continuous-time.
Assume that, for any $r$, there exists $C_r$ such that
$$
\mathcal{K}(x,x) < C_r<\infty, \quad \forall x \in B_{\infty}^r. 
$$
Then, the RKHS $\mathcal{H}$ is stable.
\end{proposition}
{\bf{Proof:}} 
Let  the system input $u \in \ell_{\infty}$ in discrete-time or
$u  \in \mathcal{L}_{\infty}$ in continuous-time.
Then, we can find a closed ball $B_{\infty}^r$ 
containing, for any $t$, all the input locations $x_t$ induced by $u$ as
defined in (\ref{xsysid}-\ref{xsysid4}).
For any $g \in \mathcal{H}$ and $x_t \in B_{\infty}^r$, 
exploiting the reproducing property and the Cauchy-Schwartz inequality, one obtains
\begin{subequations}
\begin{align*}
|g(x_t)| &= | \langle g, \mathcal{K}_{x_t} \rangle_{\mathcal{H}}|  \leq \| g \|_{\mathcal{H}}  \| \mathcal{K}_{x_t} \|_{\mathcal{H}} \\
&= \| g \|_{\mathcal{H}}  \sqrt{\mathcal{K}(x_t,x_t)} \leq  \| g \|_{\mathcal{H}} \sqrt{C_r}, 
\end{align*}
\end{subequations}
hence proving the stability of $\mathcal{H}$.  
\begin{flushright}
$\blacksquare$
\end{flushright}

The following result will be also useful later on.
It derives from the fact that 
kernels products (sums) induce RKHSs of functions 
which are products (sums) of the functions 
induced by the single kernels \cite{Aronszajn50}[p. 353 and 361].

\begin{proposition}{\bf{(RKHS stability with kernels sum and product)}} \label{StableNonLinear2}
Let $\mathcal{K}_1$ and $\mathcal{K}_2$ be stable kernels. Then, the RKHSs induced by 
 $\mathcal{K}_1 \times \mathcal{K}_2$ and $\mathcal{K}_1+\mathcal{K}_2$
 are both stable.
\end{proposition}

The two propositions above  allow to easily prove  
the stability of a very large class of kernels, as
discussed in discrete-time
in the remaining part of this section. 

\paragraph*{Radial basis kernels}

First, consider the  
input locations (\ref{xsysid}) contained in
$\mathcal{X} \subseteq \mathbb{R}^m$. 
As already mentioned in Section \ref{StableLinKer}, radial basis kernels  
$\mathcal{K}(x,a) = h(| x-a |)$, with $| \cdot |$ now to indicate the Euclidean norm, 
are widely adopted in machine learning. 
Important examples are the   
\emph{Gaussian kernel}
\begin{equation}\label{GK}
\mathcal{K}(x,a) = \exp\left(-\frac{ |x-a |^2}{\eta}\right), \quad  \eta>0 
\end{equation}
and the \emph{Laplacian kernel}
\begin{equation}\label{LK}
\mathcal{K}(x,a) = \exp\left(-\frac{| x-a |}{\eta} \right), \quad  \eta>0. 
\end{equation} 
From Proposition \ref{StableNonLinear}  one immediately sees 
that both these kernels are stable.
More in general, all the radial basis kernels  
are stable\footnote{This statement should not be confused with the result 
discussed in Section \ref{StableLinKer} in the linear system scenario. 
There, we have seen that radial basis kernels 
lead to unstable linear kernels $\mathcal{K}$
when used to define $K$ in the IIR (\ref{IIRker}) and
continuous-time (\ref{CTker}) case. 
Here, the Gaussian and Laplace kernels are instead used 
to define directly $\mathcal{K}$ in the nonlinear system scenario.} since they are constant 
along their diagonal ($\mathcal{K}(x,x)=h(0)$).\\
However, despite their stability,  some drawbacks 
affect the use of (\ref{GK},\ref{LK}) in system identification. 
First, the fact that $\mathcal{K}(x,x)$ is constant implies that 
these models do not include the information that output energy
is likely to increase if input energy augments.
Second, they measure the similarity among input locations 
without using the information that
$u_{t-\tau}$ is expected to have less influence on the prediction of $y_t$
as the positive lag $\tau$ augments.
Such limitation is also in some sense hidden  
by the finite-dimensional context. 
In fact, if the input locations are now defined by 
(\ref{xsysid2}), i.e. the system memory is infinite,
the Gaussian kernel becomes
\begin{equation}\label{GK2}
\mathcal{K}(x,a) = \exp\left(-\frac{\|x-a\|_2}{\eta}\right).
\end{equation}
This model is not reasonable:
it is not continuous  around the origin of $\mathbb{R}^{\infty}$
and, out of the diagonal, is null for many input locations.
This reveals the importance of
finding an appropriate metric to measure the distance between different input trajectories. 

{\bf New kernel for nonlinear system identification}   
We now show how stable spline kernels,
which embed exponential stability of linear systems \cite{SS2010},
can be useful also to define nonlinear models. 
Specifically, let $K_{\alpha}$
be a stable spline kernel, e.g. diagonal with
$K_{\alpha}(i,i)=\alpha^{\max(i,j)}$ or given by
$K_{\alpha}(i,j)=\alpha^{\max(i,j)}$ for any integer $i$ and $j$. 
Then, 
define the \emph{nonlinear stable spline} (NSS) kernel as
\begin{equation}\label{KerNonLin1}
\mathcal{K}(a,x) = a^T K_{\alpha} x \times \exp\left(-\frac{ (a-x)^T K_{\alpha} (a-x)}{\eta}\right)   \quad  \ \  \text{NSS},
\end{equation}
which corresponds to 
the product between a linear kernel 
and a modified version of (\ref{GK}). 
Such kernel 
defines a new infinite-dimensional RKHS suited for identification of nonlinear output error models. 
Being no more constant along the diagonal, it embeds the information 
that output energy augments if input energy increases, preserving BIBO stability 
(as one can easily deduce from Propositions \ref{StableNonLinear} and \ref{StableNonLinear2}). 
Note also that, letting the dimensionality $m$ of the
regression space go to infinity, the  
difficult selection of the discrete dimension of the regressors $x$
has been eliminated. In fact, input locations similarity is regulated by the hyperparameter $\alpha$
that includes the information that the influence of $u_{t-\tau}$ on $y_t$ goes  to zero
as the time lag $\tau$ increases.

\subsection{Numerical experiment}

The following nonlinear system is taken from
\cite{Spinelli2005}:
\begin{eqnarray*} 
\tiny
f(x_t) &=&u_{t}+0.6u_{t-1}+0.35\left(u_{t.2}+u_{t-4} \right)-0.25u^2_{t-3}\\
&+&0.2(u_{t-5}+u_{t-6})+0.9u_{t-3}+0.25u_{t}u_{t-1}+0.75u^3_{t-2}\\
&-&u_{t-1}u_{t-2}+0.5\left(u^2_{t}+u_{t}u_{t-2}+u_{t-1}u_{t-3}\right) 
\end{eqnarray*}
\noindent Then, consider the identification of the following two systems, called (S1) and (S2):
$$
y_t = f(x_t) + e_t \ \ \ \ \text{(S1)}, \qquad y_t = \sum_{k=1}^\infty \theta_k u_{t-k} + f(x_t) + e_t  \ \ \  \ \text{(S2)}, 
$$
%
%
\noindent where all the $u_t$ and $e_t$ are independent Gaussian noises of variance 4.
Note that (S2) contains the sum of a linear time invariant system 
(details on the impulse response $\theta$ are given below) and the nonlinear FIR in (S1).
Our aim is to identify the two systems from 1000 input-output pairs $(x_t,y_t)$ via
(\ref{RN}). The performance will be measured by the percentage fit on a test set of 1000 noiseless system outputs 
contained in the vector $y^{test}$, i.e. 
\begin{equation}\label{fit}
100 \% \left(1-\frac{|y^{test}- \hat{y}^{test}|}{|y^{test}-\bar{y}^{test}|}\right), 
\end{equation}
where $\bar{y}^{test}$ is the mean of the components of $y^{test}$ while $\hat{y}^{test}$ is the prediction from an estimated model. 
We will display MATLAB boxplots of the 100 fits achieved by (\ref{RN}) after a Monte Carlo of 100 runs, using different kernels.
At any run, independent noises are drawn to form new identification and test data. 
The impulse response $h$ in (S2) also varies. It is a $10$-th order rational transfer function with $\ell_2$ norm 
equal to 10 (this makes similar the contribution to the output variance of the linear and nonlinear system components) and poles inside the complex circle of radius 0.95, randomly generated 
as detailed in \cite{Pillonetto2016}[section 7.4].\\
First, we use the Gaussian kernel (\ref{GK}) over an $m$-dimensional input space.
Plugged in  (\ref{RN}), it defines the hyperparameter vector $[m \ \eta \ \gamma]$.
For tuning the Gaussian kernel hyperparameters,
the regressor vector dimension $m$ is chosen by an oracle not implementable in practice.
Specifically, at any run, for each $m \in \{1,\ldots,50\}$ the pair $(\eta,\gamma)$ is determined 
via marginal likelihood optimization \cite{DeNicolao1} using only the identification data.
Multiple starting points have been adopted to mitigate the effect of local minima. 
The oracle has then access to the test set to select, among the 50 couples,
that maximizing the prediction fit (\ref{fit}).
\begin{figure}
		\begin{center} 
		\ 
			\includegraphics[width=0.3\textwidth]{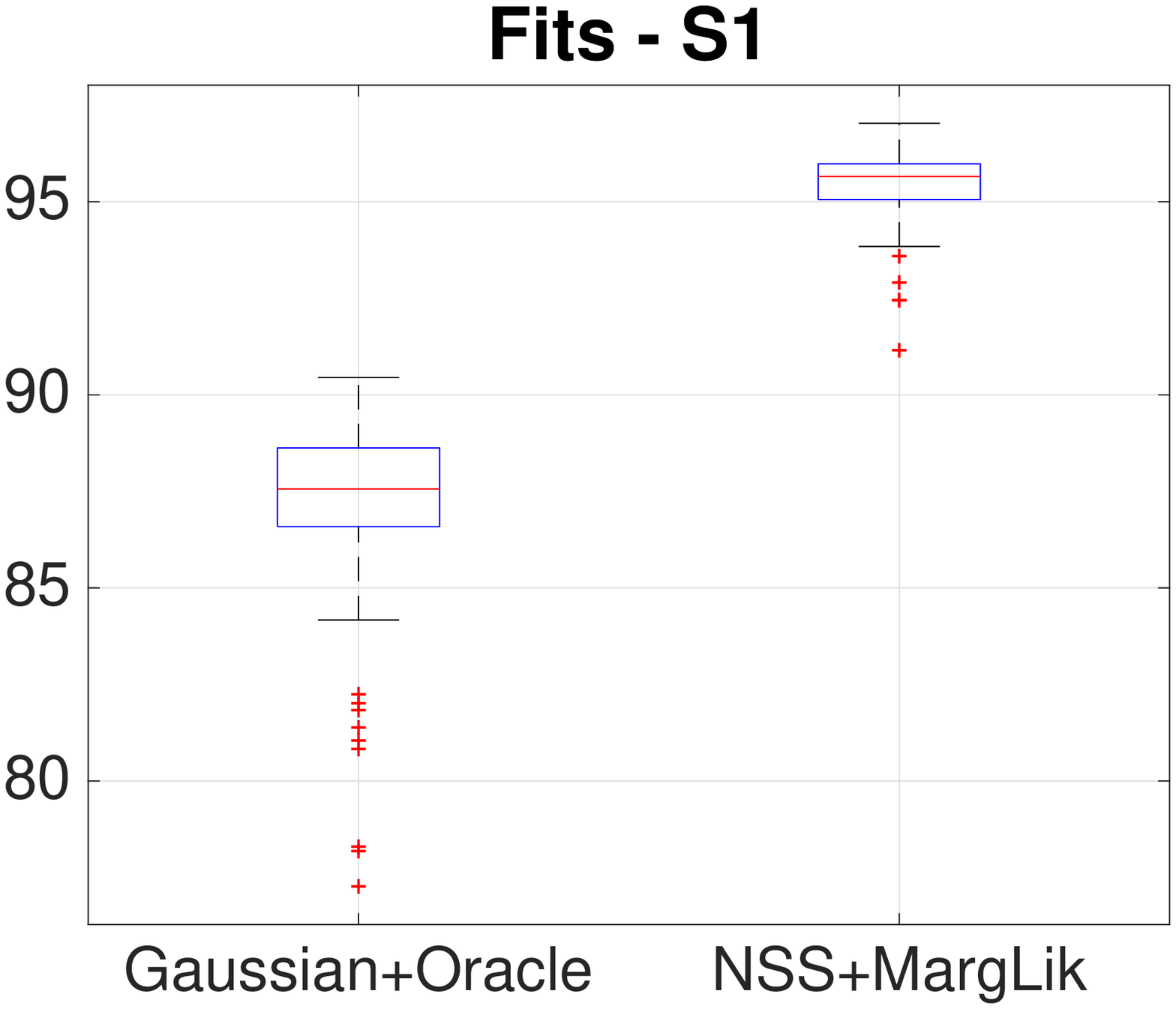} \\ 
			 \includegraphics[width=0.3\textwidth]{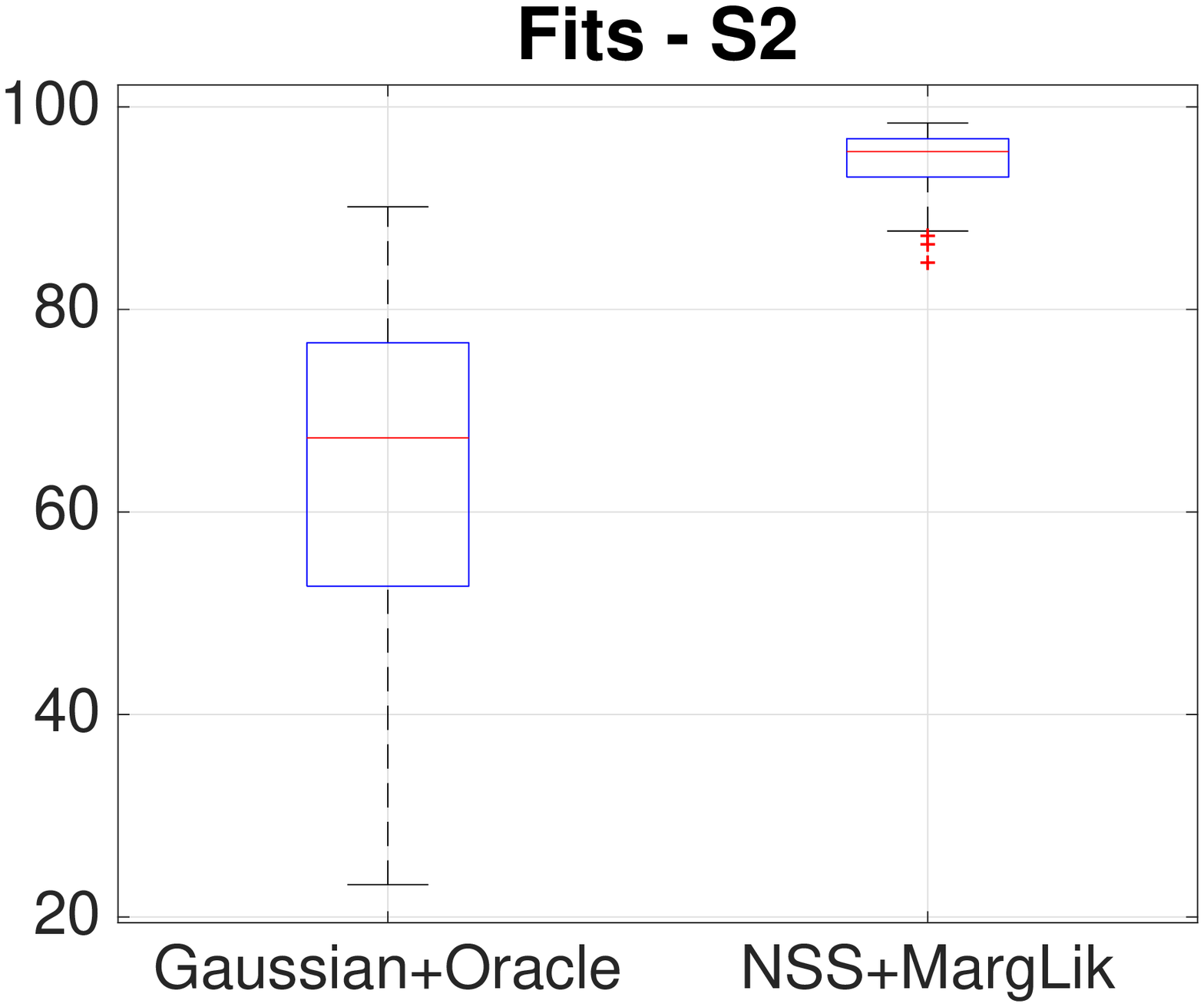} 
		\end{center}
		\caption{Boxplots of the 100 test set fits achieved in the two scenaria by (\ref{RN}) equipped with the Gaussian kernel (\ref{GK}) relying on the oracle
		to select the regressors space dimension (left boxplots) and by the new kernel NSS (\ref{KerNonLin1})
		with all the hyperparameters tuned via marginal likelihood optimization (right boxplots).}
\label{Fig1}
\end{figure}
\noindent The two left boxplots in Fig. \ref{Fig1} report the prediction fits achieved by this  procedure  
applied to identify the first (top) and the second (bottom) system. 
Even if the Gaussian kernel is equipped with the oracle, its performance is satisfactory only 
in the (S1) scenario while the capability of predicting outputs in the (S2) case is poor
during many runs.  In place of the marginal likelihood, a cross-validation strategy has been also used for tuning  $(\eta,\gamma)$,
obtaining results similar to those here displayed.\\
During the Monte Carlo, even when the number of impulse response coefficients $\theta_k$ different from zero is quite large,
we have noticed that a relatively small value for $m$ is frequently chosen, i.e. the oracle tends to use few past input values $(u_{t-1},u_{t-2},\ldots)$  to predict $y_t$. This indicates that the Gaussian kernel structure induces the oracle to introduce a significant bias to guard the estimator's variance.\\ 
%
Now, we show that in this example model complexity can be better controlled 
avoiding the difficult and computationally expensive choice of discrete orders.
In particular, we set $m=\infty$ and use 
the new NSS kernel (\ref{KerNonLin1}).  
For tuning the NSS hyperparameter vector $(\alpha,\eta,\gamma)$, 
no oracle having access to the test set is employed but just  
a single continuous optimization of the marginal likelihood that uses only the identification data.
The fits achieved by NSS after the two Monte Carlo studies are  in the right boxplots of the two figures above:
in both the cases the new estimator behaves very nicely. 


\section{Consistency of regularization networks for system identification}\label{Sec5}

\subsection{The regression function}\label{AsymptoticSub1}

In what follows, the system input $u$ is a stationary stochastic process over
$\mathbb{Z}$ in discrete-time or $\mathbb{R}$ in continuous-time. 
The distribution of $u$ induces on $\mathcal{X}$ the (Borel non degenerate) probability measure $\mu_x$,
from which the input locations $x_i$ are drawn. 
In view of their dependence on $u$, the $x_i$ are in general correlated each other 
and unbounded, e.g. for Gaussian $u$ no bounded set contains $x_i$ with probability one.
Such peculiarities, inherited by the system identification
setting, already violate the data generation assumptions routinely adopted in machine learning.\\
The identification data $\{x_i,y_i\}_{i=1}^{\infty}$ are assumed to be a stationary stochastic process.
In particular, each couple $(x,y)$ has joint probability measure  $\mu_{yx}(y,x)=\mu_{y|x}(y|x)\mu_{x}(x)$ where 
$\mu_{y|x}$ is the probability measure of the output $y$ conditional on a particular input location $x$.\\
Given a function (dynamic system) $f$, the least squares error associated to $f$ is 
\begin{equation}\label{ErrSquareLoss}
\mathcal{E}  (y-f(x))^2 = \int_{\mathcal{X} \times \mathbb{R}} \ \left(y-f(x)\right)^2 d\mu_{yx}(y,x). 
\end{equation}
The following result is well known and characterizes the minimizer of (\ref{ErrSquareLoss}) which goes under the name of
 \emph{regression function} in machine learning.
\begin{theorem}[The regression function]\label{RegFun}
We have
$$
f_{\rho} = \argmin_f \ \mathcal{E}  (y-f(x))^2 , 
$$ 
where $f_{\rho}$ is the \emph{regression function} defined for any $x \in\mathcal{X}$ by
 \begin{equation}\label{RegressionFun}
f_{\rho}(x) = \int_{\mathbb{R}} y \ d\mu_{y|x}(y|x). 
\end{equation}
\end{theorem}
In our system identification context, $f_{\rho}$
is the dynamic system associated to the optimal predictor that minimizes
the expected quadratic loss on a new output drawn from $\mu_{yx}$.

\subsection{Consistency of regularization networks for system identification}

Consider a scenario where $\mu_{y|x}$ (and possibly also $\mu_{x}$)  
is unknown and only $N$ samples $\{x_i,y_i\}_{i=1}^N$ from $\mu_{yx}$ are available.
We study the convergence of 
the RN in (\ref{RN}) to the optimal predictor $f_{\rho}$ as $N \rightarrow \infty$ 
under the input-induced norm
$$
\| f \|_{x}^2 = \int_{\mathcal{X}} \ f^2(x) d\mu_x(x).
$$
This is the norm in the classical 
Lebesgue space $\mathcal{L}_2^{\mu_x}$ 
already introduced at the end of Section \ref{Sec2}.\\
We assume that the reproducing kernel of $\mathcal{H}$ 
admits the expansion 
$$
\mathcal{K}(x,a) = \sum_{i=1}^{\infty} \ \zeta_i \rho_i(x) \rho_i(a), \ \ \zeta_i > 0 \ \forall i,
$$
with $\zeta_i$ and $\rho_i$ defined via (\ref{Lk2}) and the probability measure $\mu_x$.\\
Exploiting the regression function, the measurements process can be written as 
\begin{equation}\label{MeasMod}
y_i = f_{\rho}(x_i) + e_i,
\end{equation}
where the errors $e_i$ are zero-mean and identically distributed.
They can be correlated each other and also with the input locations $x_i$.
Given any $f \in \mathcal{H}$ and the $\ell$-th kernel eigenfunction $\rho_{\ell}$, 
we define the random variables $\{v_{\ell i}\}_{i \in \mathcal{Z}}$ by
combining $f$, $\rho_{\ell}$ and the
errors $e_i$ defined by (\ref{MeasMod}) as follows 
\begin{equation}\label{Rli}
v_{\ell i} = \left( f(x_i) + e_i \right) \rho_{\ell}(x_i).
\end{equation}
Let $\mathcal{H}$ be stable so that the $\{v_{\ell i}\}$ form a stationary process. 
In particular, 
note that each $v_{\ell i}$  is the product of the outputs from 
two stable systems: the first one, $f(x_i) + e_i$, is corrupted by noise while the second one,
$\rho_{\ell}(x_i)$, is noiseless.\\ 
Now, by summing up over $\ell$ the cross covariances of lag $k$ using kernel eigenvalues 
$\zeta_{\ell}$ as weights one obtains\footnote{As shown in Appendix, 
the $c_k$ in (\ref{Rli}) are invariant w.r.t. the particular spectral decomposition
used to obtain the pairs $(\zeta_i,\rho_i)$.}
\begin{equation} \label{Rli}
c_{k} :=  \sum_{\ell=1}^{\infty} \  \zeta_{\ell} Cov\left(v_{\ell i}, v_{\ell, i+k} \right)
\end{equation}
where $Cov(\cdot,\cdot)$ is the covariance operator. 
The next proposition shows that    
summability of the $c_k$ is key 
for consistency. 

\begin{proposition}{\bf{(RN consistency in system identification)}}\label{RNconv}
Let $\mathcal{H}$ be the RKHS with kernel  
\begin{equation}\label{Cond1}
\mathcal{K}(x,a) = \sum_{i=1}^{\infty} \ \zeta_i \rho_i(x) \rho_i(a), \ \ \zeta_i > 0 \ \forall i,
\end{equation}
where $(\zeta_i,\rho_i)$ are the eigenvalues/eigenfunctions pairs 
defined by (\ref{Lk2}) under the probability measure $\mu_x$.  
Assume that, for any $r>0$ and $f$ s.t.  $\| f\|_{\mathcal{H}} \leq r$, there exists a constant $C_r$ such that
\begin{equation}\label{Ccondition}
\sum_{k=0}^{\infty} \ |c_{k}| < C_r < \infty,
\end{equation}
with $c_{k}$ defined in  (\ref{Rli}).
Let also 
\begin{equation}\label{GammaRule}
\gamma \propto \frac{1}{N^{\alpha}}, 
\end{equation}
where $\alpha$ is any scalar in $(0,\frac{1}{2})$.
Then, if $f_{\rho} \in \mathcal{H}$
and $\hat{g}_N$ is the RN in (\ref{RN}),
as $N \rightarrow \infty$ one has 
\begin{equation}\label{Result2}
\| \hat{g}_N - f_{\rho} \|_{x}  \longrightarrow_p 0,
\end{equation}
where $\longrightarrow_p$ denotes convergence in probability. 
\end{proposition}

The fact that $\sum_{\ell=1}^{\infty} \  |\zeta_{\ell}|< \infty$ 
is already an indication that (\ref{Ccondition})  is not so hard to be satisfied.
Indeed, to the best of our knowledge, (\ref{Ccondition}) is the weakest RKHS condition
currently available which guarantees RN consistency. 
In fact, previous works, like \cite{Zou2009,Sun2010}, 
beyond considering only noises with densities of compact support use mixing conditions (which rule out infinite memory systems) with fast mixing coefficients decay.
These assumptions largely imply (\ref{Ccondition}) as it can e.g. be deduced by Lemma 2.2 in \cite{dehling1982}.\\

From (\ref{Rli}) one can see 
that (\ref{Ccondition}) essentially reduces to studying summability of the
covariances of the $v_{\ell i} = \left( f(x_i) + e_i \right) \rho_{\ell}(x_i)$.
In particular, the covariance of such product sequences depends on the first four  
moments of the component sequences, see eq. 3.1 in \cite{Wecker1978}.
The stability of $\mathcal{H}$ is thus crucial 
to ensure the existence of the moments of the $v_{\ell i}$. 
This is e.g. connected with the use
of kernels like (\ref{IIRker},\ref{CTker},\ref{KerNonLin1}) where 
the influence of past input locations on the output decays exponentially to zero as time progresses.\\

%

The relevance and usefulness of (\ref{Ccondition}) further emerges 
if more specific experimental conditions are considered. An example is given 
below by specializing Proposition \ref{RNconv} 
to the continuous-time linear setting.
Here, the aim is to reconstruct the continuous-time impulse response of a linear system fed with a stationary input process
from a sampled and noisy version of the output. Below, one can e.g. think of the sampling instants
as $t_i= i \Delta+\delta_i$ where $\delta_i$ are identically distributed random variables with support on $[0,\Delta]$. 
Then, it is shown that, for Gaussian $u$, a  
weak condition on the  input covariance's decay rate already guarantees consistency.
This outcome can also be seen as a non trivial  
extension to the dynamic context of studies 
on functional linear regression like e.g. that illustrated in \cite{Yuan2010} under independent 
data assumptions.


\begin{proposition}{\bf{(RN consistency in continuous-time linear system identification)}}\label{RNconv2}
Let $\mathcal{H}$ be the RKHS with kernel $\mathcal{K}$ defined by the continuous-time stable spline kernel 
$K(t,s) = e^{-\beta \max(t,s)}$ with support restricted to any compact set of $\mathbb{R}_+ \times \mathbb{R}_+$. 
Assume that 
\begin{itemize}
\item the regression function $f_{\rho}$ is a continuous-time and  time-invariant linear system
with impulse response $\theta$ 
satisfying $\int \ \dot{\theta}^2(t) dt< \infty$;
\item the system input $u$ is a stationary Gaussian process 
with $Cov(u(t+\tau),u(t))$
decaying to zero as $1/\tau^{1+\delta}$ for $\delta>0$;
\item the errors $e_i$ in (\ref{MeasMod}) are white, independent of the system input.
\end{itemize} 
Then, letting $\gamma$ satisfy (\ref{GammaRule}), 
the RN in (\ref{RN}) 
is consistent, i.e.
\begin{equation}\label{Result3}
\| \hat{g}_N - f_{\rho} \|_{x}  \longrightarrow_p 0.
\end{equation}
\end{proposition}



\begin{remark}[Convergence to the true impulse response] 
The optimal predictor $f_{\rho}$ is a dynamic system, unique 
as map $\mathcal{X} \rightarrow \mathbb{R}$. 
However, considering e.g. the linear scenario in Proposition \ref{RNconv2},
such functional can be associated to different impulse responses (as
illustrated in Section \ref{KvsP}
by discussing the relationship between the space $\mathcal{H}$ of linear dynamic systems and
the space $\mathcal{I}$ of impulse responses).
Convergence to $\theta$ can be guaranteed under persistently exciting conditions related to $u$ and the 
sampling instants $t_i$ \cite{Ljung:99}.
One can then wonder which kind of impulse response estimate is obtained if such conditions are not satisfied.
The answer is obtained by \cite{SurveyKBsysid}[Theorem 3 on p. 371]
which reveals that the impulse response estimate (\ref{CTsol2}) 
associated to $\hat{g}^{N}$ coincides with (\ref{RN2}).
Thus, among all the possible impulse responses
defining the optimal predictor $f_{\rho}$, the estimator (\ref{CTsol2})
will asymptotically privilege that of minimum norm in $\mathcal{I}$.
\end{remark}

\section{Conclusions}

We have introduced a new look at system identification in a RKHS framework. 
Our approach uses RKHSs whose elements are functionals associated to dynamic systems. 
This perspective establishes a solid link between system identification and machine learning,
with focus on the problem of \emph{learning from examples}.\\ 
Such framework has led to simple derivations
of RKHSs stability conditions in both linear and nonlinear
scenarios. It has been also shown that  
stable spline kernels can be used as basic building blocks to define 
other models for nonlinear system identification.\\
In the last part of the paper, RN convergence to the optimal predictor 
has been proved under assumptions 
tailored to system identification, also pointing out 
the link between consistency
and RKHS stability. 
This  general  treatment will hopefully pave the way for an even more fruitful interplay between
RKHS theory and regularized system identification.\\ 

\section*{Appendix}

\subsection*{Proof of Proposition \ref{RNconv}}

We start reporting three useful lemmas instrumental to the main proof.
First, define  
\begin{equation}\label{hatFnoiseless}
\hat{f} = \arg \min_{f \in \mathcal{H}}  \| f-f_{\rho} \|^2_{x}+ \gamma \| f \|^2_{\mathcal{H}}
\end{equation}
and
\begin{equation}\label{Etai}
\eta_i (\cdot) = \left[y_i-\hat{f}(x_i) \right] \mathcal{K}(x_i, \cdot). 
\end{equation}
In addition, the notation $S_x: \mathcal{H} \rightarrow \mathbb{R}^N$
is the sampling operator defined by $S_x f=[f(x_1) \ldots f(x_n)]$ while $S_x^*$ is its adjoint given by
\begin{equation}\label{Sprime}
S_x^* c = \sum_{i=1}^N \ c_i \mathcal{K}_{x_i} \quad \forall \ c \in \mathbb{R}^N.
\end{equation}

The first lemma below
involves the definitions of
$\eta_i,\hat{f},S_x$ and $S_x^*$ given above.
It is derived from \cite{Smale2007} and Proposition 1 in \cite{Smale2005}.

\begin{lemma}\label{LemmaTwoBounds}
It holds that
\begin{equation}\label{FirstUs1}
\mathcal{E} \eta_i = \gamma \hat{f}. 
\end{equation}
Furthermore, if $v \in \mathcal{H}$ and $f$ satisfies
$$
 \left( \frac{S_x^* S_x}{N} + \gamma I\right) f = v,
$$
where $I$ denotes the identity operator, one has 
 \begin{equation}\label{FirstUs3}
\|f \|_{\mathcal{H}} \leq \frac{1}{ \gamma}\|v \|_{\mathcal{H}}.
\end{equation}
\end{lemma}

The second lemma states a bound between the expected RKHS distance between
$\hat{g}_N$ and $\hat{f}$. 

\begin{lemma}\label{LemmaM2}
Let $r= 2\| f_{\rho} \|_{\mathcal{H}}$. Then, for any $\gamma>0$ one has
\begin{equation}\label{IneqCr}
\mathcal{E} \mathcal \| \hat{g}_N-\hat{f} \|_{x} \leq \frac{1}{\gamma} \sqrt{\max\left(1,\max_i \ \zeta_i \right)}  \sqrt{\frac{2C_r}{N }}.
\end{equation}
\end{lemma} 
{\bf{Proof:}}
It comes from the representer theorem and (\ref{Sprime}) that
$$
\hat{g}_N=S_x^* \left(\mathbf{K}+N \gamma I_{N}\right)^{-1}Y.
$$
Then, we have
\begin{eqnarray*} 
&&  \left( \frac{S_x^* S_x}{N} + \gamma I \right) \hat{g}_N \\
 \qquad &=&  \frac{S_x^*}{N} \left(   \mathbf{K} \left(\mathbf{K}+N \gamma I_{N}\right)^{-1}  + N \gamma  \left(\mathbf{K}+N \gamma I_{N}\right)^{-1}\right) Y\\
 \qquad &=& \frac{S_x^*}{N}Y.
\end{eqnarray*}
Hence, one has
\begin{eqnarray*}
\hat{g}_N - \hat{f}  &=&  \left( \frac{S_x^* S_x}{N} + \gamma I\right)^{-1} \left(\frac{S_x^* Y}{N}- \frac{S_x^* S_x \hat{f}}{N}-\gamma \hat{f} \right)\\
 &=&  \left( \frac{S_x^* S_x}{N} + \gamma I\right)^{-1}   \frac{1}{N}  \sum_{i=1}^{N}  \left ( \eta_i - \mathcal{E}[\eta_i ] \right ),
\end{eqnarray*}
where we used 
the equality $S_x^* Y - S_x^* S_x \hat{f} = \sum_{i=1}^{N} \eta_i$ and (\ref{FirstUs1}). 
Using (\ref{FirstUs3}) in Lemma \ref{LemmaTwoBounds}, we then obtain
\begin{equation}\label{InvGamma}
 \| \hat{g}_N-\hat{f} \|_{\mathcal{H}} \leq \frac{1}{\gamma} \left\| \frac{1}{N}  \sum_{i=1}^{N}  \left ( \eta_i - \mathcal{E}[\eta_i ] \right )   \right\|_{\mathcal{H}}.
\end{equation}
Now, let $f :=f_{\rho}-\hat{f}$. 
With the $e_i$ defined in (\ref{MeasMod}), we can write 
\begin{eqnarray*}
&& \eta_i (\cdot) -\mathcal{E} \eta_i (\cdot)\\
&=& \left[ f(x_i) + e_i \right] \mathcal{K}(x_i, \cdot) - \mathcal{E}  \left[ f(x_i) + e_i \right] \mathcal{K}(x_i, \cdot) \\
&=&    \sum_{\ell=1}^{\infty}   \zeta_{\ell} \left[ (f(x_i) + e_i)\rho_{\ell}(x_i) - \mathcal{E}(f(x_i) + e_i)\rho_{\ell}(x_i) \right]  \rho_{\ell}(\cdot)\\ 
&=&    \sum_{\ell=1}^{\infty}  \zeta_{\ell}  \left[  v_{\ell i}  -  m_{\ell}  \right]   \rho_{\ell}(\cdot)
\end{eqnarray*}
where $v_{\ell i} = \left( f(x_i) + e_i \right) \rho_{\ell}(x_i)$ and $m_{\ell} = \mathcal{E} v_{\ell i}$.
Now, the structure of the  RKHS norm outlined in (\ref{Hrep5}) allows us to write 
\begin{eqnarray*}
&& \langle \eta_i (\cdot) -\mathcal{E} \eta_i,  \eta_j (\cdot) -\mathcal{E} \eta_j  \rangle_{\mathcal{H}} \\
&=& \langle \sum_{\ell=1}^{\infty}   \zeta_{\ell}  \left(  v_{\ell i}  -  m_{\ell}  \right) \rho_{\ell}(\cdot),  \sum_{\ell=1}^{\infty}   \zeta_{\ell}  \left(  v_{\ell j}  -  m_{\ell}  \right)  \rho_{\ell}(\cdot) \rangle_{\mathcal{H}} \\
&=& \sum_{\ell=1}^{\infty} \zeta_{\ell} \left(v_{\ell i}  -  m_{\ell} \right) \left(v_{\ell j}  -  m_{\ell} \right).
\end{eqnarray*}
So, using definition (\ref{Rli})
\begin{eqnarray*}
&& \mathcal{E} \langle \eta_i  -\mathcal{E} \eta_i ,  \eta_j -\mathcal{E} \eta_j  \rangle_{\mathcal{H}}\\ 
\qquad &=&\sum_{\ell=1}^{\infty}  \zeta_{\ell} Cov\left(v_{\ell i}, v_{\ell, j} \right) \\
\qquad  &=& c_{|i-j|}.
\end{eqnarray*}
Comparing the values of the objective in (\ref{hatFnoiseless}) at the optimum $\hat{f}$
and at $f_{\rho}$, one finds $\|\hat{f} \|_{\mathcal{H}} \leq \| f_{\rho} \|_{\mathcal{H}} $ so that
$$
\| f_{\rho}-\hat{f}  \|_{\mathcal{H}} = \| f \|_{\mathcal{H}} \leq 2 \| f_{\rho} \|_{\mathcal{H}}.  
$$
This, combined with (\ref{Ccondition}), implies that for any $\gamma>0$ 
$$
\sum_{k=0}^{\infty} \ |c_{k}| < C_r < \infty, \quad r= 2 \| f_{\rho} \|_{\mathcal{H}}.
$$
Hence, we obtain 
\begin{eqnarray*}
&& \mathcal{E}  \left[  \left\| \frac{1}{N}  \sum_{i=1}^{N}  \left ( \eta_i - \mathcal{E}[\eta_i ] \right )   \right\|_{\mathcal{H}}^2 \right]  \\
&& \qquad \leq \frac{1}{N^2}   \sum_{i=1}^{N} \sum_{j=1}^{N} |c_{|i-j|}| \leq  \frac{2C_r}{N}. 
\end{eqnarray*}  
Now, recall that, if $f = \sum_{i=1}^{\infty} a_i \rho_i$, then 
$\| f\|_{\mathcal{H}}^2 =  \sum_{i=1}^{\infty} \frac{a_i^2}{\zeta_i}$ while 
$\| f\|_{x}^2 =  \sum_{i=1}^{\infty} a_i^2$. Thus, 
for any $f \in \mathcal{H}$, one has
$$
\| f\|^2_{x} \leq \max\left(1,\max_i \ \zeta_i\right)  \| f\|^2_{\mathcal{H}}.
$$
The use of Jensen's inequality and (\ref{InvGamma}) then completes the proof.
\begin{flushright}
$\blacksquare$
\end{flushright}

Now, we need to set up some additional notation.
Following \cite{Smale2005,Smale2007},
given the integral operator 
\begin{equation*}
L_{\mathcal{K}}[\rho_i] := \int_{\mathcal{X}} \mathcal{K}(\cdot,u) \rho_i(u) d\mu_x(u) = \zeta_i \rho_i, 
\end{equation*} 
for $r>0$ we define 
\begin{equation*}
L_{\mathcal{K}}^{-r}[f] = \sum_{i=1}^{\infty}  \frac{c_i}{\zeta_i^r}  \rho_i.
\end{equation*}
The third lemma reported below 
contains the inequality (\ref{Bound1}) 
which was also derived in \cite{Smale2005}
assuming a compact input space. 
However, the bound holds just assuming the
validity of the kernel expansion (\ref{Cond1}).
To see this, recalling Theorem \ref{SpectralRKHS3}, first note that 
$L_{\mathcal{K}}^{-r} f_{\rho} \in \mathcal{L}_2^{\mu_x}$ for any $0 \leq r \leq 1/2$.
So, there exists $g \in \mathcal{L}_2^{\mu_x}$, say $g=\sum_{i=1}^{\infty} d_i \rho_i$,
such that $f_{\rho}=\sum_{i=1}^{\infty} \zeta_i^r d_i \rho_i$.
After simple computations, one obtains 
$\hat{f}  - f_{\rho} = - \sum_{i=1}^{\infty} \frac{\gamma}{\zeta_i+\gamma} \zeta_i^r d_i \rho_i$
and the same manipulations contained in the proof of Theorem 4 
in  \cite{Smale2005}[p. 295] lead to the following result.

\begin{lemma}\label{LemmaM1}
For any $0 < r \leq 1/2$, one has
\begin{equation}\label{Bound1}
\| \hat{f}-f_{\rho} \|_{x} \leq \gamma^{\ r}  \| L_{\mathcal{K}}^{-r}f_{\rho} \|_{x}
\end{equation} \end{lemma}
Combining (\ref{IneqCr}) and  (\ref{Bound1}), 
for any $0 < r \leq 1/2$ it holds that
\begin{equation}\label{DecomErr2}
\mathcal{E} \| \hat{g}_N-f_{\rho} \|_{x}  \leq  \gamma^{\ r} \| L_{\mathcal{K}}^{-r} f_{\rho} \|_{x} 
+ \frac{1}{\gamma} \sqrt{\max\left(1,\max_i \ \zeta_i \right)}  \sqrt{\frac{2C_r}{N }}.  
\end{equation}
Hence, when $\gamma$ is chosen according to (\ref{GammaRule}), $\mathcal{E}\| \hat{g}_N-f_{\rho} \|_{x}$ 
converges to zero as $N$ grows to $\infty$. Using the Markov inequality, (\ref{Result2}) is finally obtained.

\subsection*{Proof of Proposition \ref{RNconv2}}
Let $\mathcal{I}$ be
the space of impulse responses with compact support e.g. on $[0,T]$
induced by the stable spline kernel $K$.
Then, it comes from  \cite{PillACC2010} that
$\| \theta\|_{\mathcal{I}}^2 \propto \int_{0}^T \dot{\theta}^2(t) e^{\beta t} dt$.
So, finite energy of the first derivative of $\theta$ ensures that  
the optimal predictor $f_{\rho}$ belongs the RKHS $\mathcal{H}$ defined by the linear kernel $\mathcal{K}$
induced by $K$.\\
Now, we have just to prove that condition
(\ref{Ccondition}) holds. 
Recall that the $c_k$ are invariant w.r.t. the particular kernel expansion of $\mathcal{K}$ adopted.
For the stable spline kernel $K$ we choose the expansion 
$K(t,\tau)= \sum_{\ell=1}^{\infty} \zeta_{\ell} \psi_{\ell}(t) \psi_{\ell}(\tau)$
derived in \cite{PillACC2010} where
$$
\psi_{\ell}(t) = \sqrt{2}\sin\left( \frac{e^{-\beta s }}{\sqrt{\zeta_{\ell}}} \right), \quad   \zeta_{\ell} = \frac{1}{(\ell \pi - \pi/2)^2}.
$$
The eigenfunctions thus satify 
\begin{equation}\label{AA1}
|\psi_{\ell}(t)|<\sqrt{2} \ \ \forall (\ell,t).
\end{equation}
Now, let $f$ be any dynamic system satisfying $\| f \|_{\mathcal{H}} \leq r$ 
and let $\theta$ be the associated impulse response of minimum norm living in $\mathcal{I}.$\footnote{
The minimum norm impulse
response is chosen without loss of generality since any other $\theta$ associated with
$f$ would induce the same
input-output relationship.}
From the arguments discussed in section \ref{KvsP} one then has $\| \theta \|_{\mathcal{I}} \leq r$. 
It then holds that 
\begin{equation}\label{AA2}
|\theta(t)|=   | \langle \theta, K_t  \rangle_{\mathcal{I}}| \leq r \sqrt{K(t,t)}  \implies \max_{t \in [0,T]} |\theta(t)| \leq A_r< \infty
\end{equation}
with $A_r$ independent of the particular $f$ chosen inside the ball of radius $r$ of $\mathcal{H}$.\\
Without loss of generality, the input $u$ is now assumed zero-mean so that
the $f(x_i)$ and $ \rho_{\ell}(x_i)$ become zero-mean Gaussian processes. 
Using eq. 3.2 in \cite{Wecker1978}, for $k>1$ one obtains
\begin{eqnarray}\nonumber
Cov\left(v_{\ell i}, v_{\ell, i+k} \right) &=& Cov\left(f(x_i) \rho_{\ell}(x_i) , f(x_{i+k}) \rho_{\ell}(x_{i+k}) \right)\\\nonumber
&=& Cov\left(f(x_i),f(x_{i+k})\right) Cov\left( \rho_{\ell}(x_i), \rho_{\ell}(x_{i+k})\right)     \\\nonumber
&+& Cov\left(f(x_i),\rho_{\ell}(x_{i+k})\right) Cov\left(f(x_i),\rho_{\ell}(x_{i-k})\right).\\
\label{4cov}
\end{eqnarray}
Now, let $h$ be any of the four covariances in the r.h.s. of (\ref{4cov}). 
Combining (\ref{AA1},\ref{AA2}) and classical integral formulas for covariances computations, 
as e.g. reported in \cite{PapoulisProb}[p. 308-313], it is easy to obtain  
a constant $B_r$ independent of $\ell$ such that $|h(k)| \leq B_r/ k^{1+\epsilon}$.
Condition  (\ref{Ccondition}) thus holds true and this completes the proof.

\bibliographystyle{plain}
\bibliography{biblio}
\end{document}